\documentclass[a4paper,11pt]{article}
\pdfoutput=1 % if your are submitting a pdflatex (i.e. if you have
\usepackage{jcappub} % for details on the use of the package, please
\usepackage[T1]{fontenc} % if needed
\usepackage{comment}

\usepackage{subfigure}
\usepackage{color}
\usepackage{multirow}
\usepackage{xcolor}
\usepackage{soul}
\title{\boldmath Gaussian process reconstructions and model building of quintom dark energy from latest cosmological observations}

\author[a,b,1]{Yuhang Yang, \note{Co-first author.}}
\author[a,b,c,1]{Qingqing Wang,}
\author[a,b]{Chunyu Li,}
\author[a,b]{Peibo Yuan,}
\author[a,b,d,2]{Xin Ren, \note{Corresponding author.}}
\author[e,b,f,2]{Emmanuel N. Saridakis,}
\author[a,b,2]{Yi-Fu Cai}

\affiliation[a]{Department of Astronomy, School of  Physical Sciences, 
University of Science and Technology of China,  96 Jinzhai Road, Hefei, Anhui 
230026, China}
\affiliation[b]{CAS Key Laboratory for Research in  Galaxies and Cosmology, 
School of Astronomy and Space Science, University of Science and Technology of 
China, 96 Jinzhai Road, Hefei, Anhui 230026, China}
\affiliation[c]{Kavli IPMU (WPI), UTIAS, The University of Tokyo, Kashiwa, Chiba 277-8583, Japan}
\affiliation[d]{Department of Physics, Tokyo Institute of Technology, 2-12-1 
Ookayama, Meguro-ku,  Tokyo 152-8551, Japan}
\affiliation[e]{National Observatory of Athens, Lofos Nymfon, 11852 Athens, 
Greece}
\affiliation[f]{Departamento de Matem\'{a}ticas, Universidad Cat\'{o}lica del 
Norte, Avda. Angamos  0610, Casilla 1280 Antofagasta, Chile}

\emailAdd{yyh1024@mail.ustc.edu.cn}
\emailAdd{wangqqq@mail.ustc.edu.cn}
\emailAdd{springrain@mail.ustc.edu.cn}
\emailAdd{yuanpeibo@mail.ustc.edu.cn}
\emailAdd{rx76@ustc.edu.cn}
\emailAdd{msaridak@noa.gr}
\emailAdd{yifucai@ustc.edu.cn}

\abstract{
In this article we use the latest cosmological observations, including SNe, BAO, CC and RSD, to reconstruct the cosmological evolution via Gaussian process regression. At the background level, we find consistency with quintom dynamics for different data combinations and categorize the characteristics of dark energy into three distinct types: negative-energy dark energy, late-dominated dark energy, and oscillating dark energy. Considering the effect of modified gravity on the growth of matter perturbations, the reconstruction results at the perturbative level indicate that only minor corrections to general relativity are required. Furthermore, we provide theoretical interpretation for these three different types of dynamical dark energy behavior, within the framework of modified gravity, scalar fields, and dark energy equation-of-state parametrizations. Finally, we show that all of these models can be unified within the framework of effective field theory.
}

\begin{document}
\maketitle
\flushbottom

\section{Introduction}

The direct evidence for the Universe's current acceleration was first carried out by two groups through the observations of the luminosity distances of high-redshift supernovae in 1998 \cite{SupernovaSearchTeam:1998fmf, SupernovaCosmologyProject:1998vns}. 
Subsequently, various observations from the Cosmic Microwave Background (CMB) \cite{Planck:2018vyg}, large-scale structure (LSS) \cite{Chuang:2013hya} and other astronomical windows independently corroborated this discovery. In order to describe the behavior of the accelerated expansion, the concept of dark energy was introduced. In the standard cosmological scenario, the dark energy is expressed as the cosmological constant $\Lambda$. However, this leads to several issues such as the fine-tuning and coincidence problems in $\Lambda$CDM cosmology \cite{Weinberg:2000yb}. 
To provide a physical explanation for these problems and obtain a better understanding of the nature of dark energy, the dynamical phenomena associated with the cosmological constant were considered in \cite{Mukohyama:2003nw, Copeland:2006wr}. Among these, a class of scalar-field models was phenomenologically proposed to describe the dynamical behavior of dark energy, including quintessence \cite{Ratra:1987rm, Wetterich:1987fm}, phantom \cite{Caldwell:1999ew}, K-essence \cite{Armendariz-Picon:2000nqq, Chiba:1999ka} and so on. 
Hence, one acquires the possibility that the equation-of-state (EoS) parameters are always greater or lower than $-1$. 

In the early 21st century, high-redshift supernova data suggested the possibility of a crossing of the cosmological constant boundary (or phantom divide) \cite{Feng:2004ad, Huterer:2004ch, Elizalde:2004mq}.
This phenomenon corresponds to a class of dynamical models with an EoS evolving across $-1$, known as \textit{quintom} dark energy (a portmanteau of “quintessence” and “phantom”). 
However, the explicit construction of the quintom scenario is more challenging compared to other models due to the constraints imposed by the \textit{No-Go theorem} \cite{Cai:2009zp}. 
The implementation of the scalar field theory of dynamical dark energy should lead to a new challenge in violating the null energy condition \cite{Buniy:2005vh,Qiu:2007fd}. Theoretically, extended gravitational theories beyond general relativity (GR), can behave as a component with a dynamical EoS (see the review \cite{CANTATA:2021ktz}). Therefore, modified gravity serving as an effective form of dynamical dark energy can provide an alternative framework without considering these issues. Common models include extended curvature-based theories such as $f(R)$ gravity \cite{Starobinsky:1980te, Capozziello:2002rd, DeFelice:2010aj,Nojiri:2010wj,Nojiri:2017ncd}, torsion-based extensions of Teleparallel Equivalent of General Relativity (TEGR) such as $f(T)$ gravity \cite{Cai:2015emx, Bahamonde:2021gfp, Krssak:2018ywd, Krssak:2015oua, Aldrovandi:2013wha}, and nonmetricity-based extensions of Symmetric Teleparallel Equivalent of General Relativity (STEGR) such as $f(Q)$ gravity \cite{BeltranJimenez:2017tkd, Heisenberg:2023lru}.

With the innovation and upgrading of observational equipment, we are able to accurately measure the basic parameters of the Universe, allowing us to search for evidence of dark energy dynamics through precise observation. 
The CMB observation from the Planck satellite with additional information from the Canada France Hawaii telescope lensing survey \cite{Heymans:2012gg} pulls the constraint of the Chevallier-Polarski-Linder (CPL) model into the phantom domain and is discrepant with the standard $\Lambda$CDM cosmology at about the $2\sigma$ level \cite{Planck:2015fie, Planck:2018vyg, DES:2017myr}. 
Moreover, reconstructing the history of the evolution of dark energy without regard to specific models can have further implications. 
With the application of the non-parametric Bayesian method, $3.5\sigma$ evidence for dynamical dark energy was found by using the data from SDSS DR7, BOSS and WiggleZ \cite{Zhao:2012aw, Zhao:2017cud, Colgain:2021pmf, Pogosian:2021mcs}.
More recently, the latest measurements of baryon acoustic oscillations (BAO) from the Dark Energy Spectroscopic Instrument (DESI) showed tantalizing hints of dynamical dark energy (DESI 2024) \cite{DESI:2024mwx}. Combining with CMB and Supernova, provide $2.5\sigma$, $3.5\sigma$, and $3.9\sigma$ evidence with the dataset from PantheonPlus \cite{Scolnic:2021amr}, Union3 \cite{Rubin:2023ovl}, and DESY5 \cite{Abbott:2024agi} respectively.
The dynamical behavior of dark energy is expected to be examined precisely in the next generation of cosmological observations.
It is noted that the DESI data suggested a quintom-B behavior, where the dark energy EoS parameter transitions through $-1$, evolving from $w < -1$ to $w > -1$  over time. This deviation from the standard cosmological model motivates us to incorporate more data and employ a model-independent method to investigate the EoS parameter of dynamical dark energy and the evolutionary history of cosmic dynamics. 

In this work, we use Gaussian process regression to reconstruct the evolution of dark energy EoS parameter with different data combinations, including SNe, BAO, and cosmic chronometers (CC) data at the background level, along with the measurement of matter growth rate through redshift-space distortion (RSD) from LSS.
The manuscript is organized as follows. In Section \ref{observation}, we review the observation evidence for dynamical dark energy in light of recent observations.
In Section \ref{reconstruction}, we introduce the data sets and the method we used to reconstruct the evolution of dark energy EoS. Then, we present the results from different combinations and divide the evolution behaviors into different categories.
In Section \ref{theory}, we discuss the theoretical interpretation of different dark energy behaviors, and we provide some examples including modified gravity and field theory, and we show that all these models can be unified within the framework of effective field theory (EFT). Finally, we summarize in Section \ref{conclusion}.

\section{Dynamical dark energy in light of recent observations}\label{observation}

Since the release of DESI's first-year BAO data in April 2024 \cite{DESI:2024mwx}, dynamical dark energy has garnered significant attention. Especially when combining CMB and supernova observations, the deviation from $\Lambda$CDM within a two-parameter ($w_0w_a$CDM) dark energy model was found to be at the $3.9\sigma$ level using the Union3 SNe dataset \cite{DESI:2024mwx}. Here we provide an overview of recent researches that examine the robustness of this result and the constraints imposed by various observations on different dynamical dark energy models.

Firstly, we should mention that the behaviors of dark energy EoS can be simply classified as follows:
\begin{itemize}
   \item $w = -1$: This corresponds to the cosmological constant $\Lambda$.
   \item $w \leq -1$: The EoS lies below the cosmological constant boundary, usually called phantom dark energy.
   \item $w \geq -1$: The EoS remains above the cosmological constant boundary, usually called quintessence dark energy.
   \item $w$ crosses $-1$: The EoS is able to evolve across the cosmological constant boundary, usually called quintom dark energy.
\end{itemize}
One of the most widely used parametrizations for dynamical dark energy is the CPL parametrization. Its first-order expansion yields
\begin{equation}
w(a)=w_{0}+w_{a} \left(1-\frac{a}{a_{0}} \right),
\end{equation}
which is also known as $w_{0}w_{a}$ model \cite{Chevallier:2000qy,Linder:2002et}. 
Here we summarize the results of various observational analyses conducted within the $w_{0}w_{a}$CDM model following the release of DESI data:
\begin{itemize}
\item \textbf{BAO}: Currently, the most powerful evidence for dynamical dark energy comes from the DESI BAO data, although a part of the community is skeptical about this result.
The data point obtained from luminous red galaxies at the effective redshift $0.51$ has a noticeable deviation from the previous BAO observations. The self-consistency of DESI data was studied in \cite{Wang:2024rjd}, which found that the data point at $z=0.51$ can hardly affect the joint constraint.
At the same time, a model-independent null test used in \cite{Dinda:2024kjf} detected the individual deviations at redshift $0.51$, thus finding no strong evidence for dynamical dark energy.
Consequently, some tried to consider excluding the new data and found that the preference for dynamical dark energy over the cosmological constant was reduced to nearly $2\sigma$ \cite{Park:2024vrw}.

\item \textbf{CMB}: In order to confirm the contribution of dynamical dark energy from CMB observations, \cite{Park:2024vrw} used both CMB and non-CMB data sets to constrain the $w_{0}w_{a}$CDM model. Their findings indicate that incorporating CMB data shifts the parameter space towards the quintom domain.
In addition, other CMB experiments like the Atacama Cosmology Telescope and the South Pole Telescope also provide temperature, polarization and lensing spectra at small scales. The combination of these small-scale observations with Planck or WMAP at large angular scales showed that CMB experiments other than Planck generally weaken the evidence for dynamical dark energy. Specifically, the information that strengthened the shift towards dynamical dark energy arises from the temperature and E-mode polarization anisotropy measurements at large angular scales $l\leq30$ \cite{Giare:2024ocw}.

\item \textbf{SNe}: In a study of the $w_0 w_a$CDM cosmological model \cite{RoyChoudhury:2024wri}, it was found that without incorporating SNe data, the EoS parameter is poorly constrained. However, when combining CMB, BAO and PantheonPlus datasets, the inclusion of the cosmological constant falls within $2\sigma$. In contrast, the combination of the CMB, BAO and DESY5 data excludes the cosmological constant at more than $2\sigma$. Consequently, these findings suggest that dynamical dark energy does not yet present a robust cosmological signature. Moreover, \cite{Efstathiou:2024xcq} compared the SNe common to the DESY5 and PantheonPlus compilations, finding evidence of an offset of $0.04$ magnitudes between low and high redshifts. After correcting for this offset, the DESY5 sample shows excellent agreement with the $\Lambda$CDM cosmology. Further discussions can also be found in \cite{Chen:2024vuf}.
\end{itemize}

In addition to the $w_{0}w_{a}$ model, many other mechanisms that can implement dynamical dark energy have also received extensive attention. Some models realize dynamical dark energy by introducing additional scalar fields. For example, it was found in \cite{Gialamas:2024lyw} that the physical quintessence models fit the observation data well, and the deviations from the constant dark energy are driven mainly by low-z supernova data. Other recent research on the quintessence scenario can be found in \cite{Bhattacharya:2024hep,Thompson:2024nxf,Kadam:2024vlw,Andriot:2024jsh,Wetterich:2024ieb,Wolf:2024eph}.
Similarly, the K-essence scalar field scenario and other field-level models were studied in \cite{Hussain:2024qrd,SpurioMancini:2024qic}.

Other works also consider different parametrizations of the dark energy EoS.
In \cite{Carloni:2024zpl}, the authors considered three typologies of models to realize cosmic acceleration, which were related to thermodynamics, Taylor expansions of the barotropic factor, and ad hoc dark energy parameterizations. They found that the best model to fit the DESI BAO, OHD and the Pantheon SNe data is a complicated log-corrected provided by the Anton-Schmidt dark energy EoS. 
Similarly, \cite{DESI:2024kob} explored three physically motivated behaviors of dark energy EoS and energy density, including the thawing class (matching many simple quintessence potentials), the emergent class (where dark energy arises recently, as in phase transition models), and the mirage class (where phenomenologically the distance to CMB last scattering is close to that from a cosmological constant $\Lambda$), and they found that the mirage class behaves essentially as well as $w_{0}w_{a}$CDM although having one fewer parameters. Furthermore, \cite{Roy:2024kni} introduced an alternative two-parameter parameterization of the dark energy EoS, which can be approximated to the CPL form at high redshifts. Their results consist with \cite{DESI:2024mwx}, and this model reduces the Hubble tension to about $2.8\sigma$ compared to the data of the Hubble Space Telescope and SH0ES and to $1.6\sigma$ with the standardized TRGB and supernova data.
Other similar research is available in \cite{Orchard:2024bve, Reboucas:2024smm,Giare:2024gpk,Dhawan:2024gqy}.

In addition to choosing a specific model, one can also choose other methods to directly reconstruct $w(z)$ and aquire intuition for the evolution of dark energy.  In the following, we summarize the recent research on the reconstruction of dynamical dark energy.

Before the release of the DESI 2024 data, the Bayesian method was used to reconstruct dynamical dark energy. Evidence at $3.5\sigma$ was found using BAO data from SDSS DR7, BOSS, and WiggleZ 
\cite{Zhao:2012aw, Zhao:2017cud, Colgain:2021pmf, Pogosian:2021mcs}. Meanwhile, Gaussian process regression was also used to reconstruct dark energy EoS with the combination of PantheonPlus + CMB + CC \cite{Wang:2022xdw} and JLA + H(z) + CMB + HII + GRB\cite{Wang:2019ufm}, and obtained a quintessence-like EoS and a quintom-like EoS at the $2 \sigma$ confidence respectively. Other studies related to Gaussian process regression can be seen \cite{Ren:2021tfi, Ren:2022aeo, Perenon:2022fgw, Zhang:2018gjb,Yang:2024tkw}.
Additionally, one can derive the evolution of $w$ by solving the characterization function, yielding results consistent with predictions from the quintom model \cite{Wang:2024qan}. 
Moreover, several works have utilized artificial neural networks \cite{Dialektopoulos:2023jam}, Ridge Regression approaches \cite{Huang:2020xyz}, and model-independent joint analyses \cite{Bonilla:2020wbn} to investigate the dynamical evolution of dark energy.

After the release of the DESI BAO data, \cite{Yang:2024kdo} reconstructed the dynamical dark energy using Gaussian process regression based on DESI as well as previous BAO data containing SDSS and WiggleZ. Their results indicated that $w$ exhibits a quintom-B behavior, crossing $-1$ from phantom to quintessence regime at $z\simeq2.18$.
A binned parameterization method was used in \cite{Pang:2024qyh}, which considers a constant EoS in three redshift bins to reconstruct the EoS based on DESI BAO + Planck 2018 CMB + SNe from Pantheon/PantheonPlus/Union3. They found that the values of $w$ lie in the quintessence regime at low redshift ($ 0 \leq z <0.4$) with $1.9\sigma$/$2.6\sigma$/$3.3\sigma$ confidence level, while lying in the phantom regime at high redshift ($ 0.8 \leq z < 2.1$) with $1.6\sigma$/$1.5\sigma$/$1.5\sigma$ confidence level, in these three data combinations with different choices of type Ia supernova datasets, respectively. 
The implemented crossing statistics method was used to reconstruct dark energy by using the data: DESI BAO only, DESI BAO+Union3, BAO+Union3+Plank. All of them show quintom-B behavior, and DESI BAO only yields a higher $w$ in the late universe \cite{DESI:2024aqx}.
In addition, the Chebyshev reconstruction has been studied by \cite{Liu:2024gfy} with DESI BAO data, DESI BAO+PantheonPlus and DESI BAO+PantheonPlus+CMB. After removing LRG1 ($z_{\mathrm{eff}}=0.51$) and LRG2 ($z_{\mathrm{eff}}=0.71$), they found that the DESI BAO only preferred phantom behavior while others had a trend of $\Lambda$CDM compared to the previous results.
Similar analyses can be found in \cite{Reboucas:2024smm, Dinda:2024ktd, Ishak:2024jhs, Zheng:2024qzi,Wolf:2024stt,Jiang:2024xnu}.
In summary, the strongest current evidence for dynamical dark energy  
comes from the joint constraints of DESI BAO and DESY5 SNe. However, claims regarding dynamical dark energy appear premature without further investigations into potential systematic effects \cite{Colgain:2024mtg}.

\section{Reconstruction}\label{reconstruction}

In this section, we provide the details of the reconstruction method. We first start from the presentation of the data that we will use, then we present the method, and finally, we discuss the obtained results.

\subsection{Data}

\textbf{SNe}-- The Type Ia SNe produces a consistent peak luminosity due to the uniform mass of white dwarfs that explode via the accretion mechanism. This characteristic allows SNe Ia to serve as standard candles for measuring distances to their host galaxies, as the observed magnitude of the supernovae predominantly depends on the distance. The relationship between the observed magnitude and the distance can be described by the following equation:
\begin{equation}
    \mu(z) \equiv m(z)-M_{\mathrm{B}}=5 \log_{10} \Bigg ( \frac{D_\mathrm{L}(z)}{10^{-5} \rm{Mpc}} \Bigg ),
\end{equation}
where $\mu$ represents the distance modulus, $m(z)$ is the standardized apparent magnitude, $M_{\mathrm{B}}$ is the B-band absolute magnitude of the standardized supernova light curve, and $D_\mathrm{L}(z)$ denotes the luminosity distance at the redshift $z$. Thus, they provide access to the luminosity distance at the redshift measured from the host galaxy. Through the relation between comoving distance $D_\mathrm{M}$ and $D_\mathrm{L}$, we obtain
\begin{equation}
    D_\mathrm{M}(z)=\frac{D_\mathrm{L}(z)}{1+z}=\frac{10^{[\mu(z)-25]/5}}{1+z},
\end{equation}
where the comoving distance in a flat universe is defined as
\begin{equation}
    D_\mathrm{M}(z)=\int_0^z\frac{c}{H(z')}\mathrm{d}z'.
\end{equation}
Then the Hubble parameter $H(z)$ can be calculated as $H(z)=c/D'_\mathrm{M}(z)$. From this relation, it becomes evident that when we set $z = 0$ the value of the absolute magnitude $M_{\mathrm{B}}$ degenerates with the value of the Hubble constant $H_0$.

For our analysis of SNe, we utilize data from three survey compilations: PantheonPlus ($0.001 < z < 2.26$) \cite{Brout:2022vxf}, Union3 ($0.05 < z < 2.26$) \cite{Rubin:2023ovl}, and DESY5 ($0.025 < z < 1.3$) \cite{Abbott:2024agi}. The values of the absolute magnitude $M_{\mathrm{B}}$ for each dataset are summarized in Table~\ref{tab:MB value}. It is also important to note that the lower values of $M_{\mathrm{B}}$ reported for the Union3 and DESY5 datasets reflect that $M_{\mathrm{B}}$ has already been incorporated into the data processing. Therefore, for these two datasets, the relevant parameter should be interpreted as $\Delta M_{\mathrm{B}}$. And it is worth to emphasize that a global, uniform shift in the absolute magnitude $M_\mathrm{B}$ of SNe Ia does not affect cosmological parameter constraints. This is because such a shift is entirely degenerate with the Hubble constant $H_0$.

\begin{table}
\centering
\caption{The mean values and associated errors of the absolute magnitude $M_{\mathrm{B}}$ (or $\Delta M_{\mathrm{B}}$ for Union3 and DESY5) used in this study, which can be found in Table~1 of \cite{Matthewson:2024ffb}.}
\begin{tabular}{cccc}
\hline
\hline
    Parameter &PantheonPlus &Union3 &DESY5 \\
\hline
    $M_{\mathrm{B}}$($\Delta M_{\mathrm{B}}$) &$-19.420$ &$-0.14894$ &$-0.057596$ \\
    $\sigma_{M_\mathrm{B}}$ &$0.155$ &$0.09129$ &$0.21322$ \\
\hline
\end{tabular}
\label{tab:MB value}
\end{table}

\textbf{BAO}-- The sound horizon $r_{\mathrm{d}}$ of the baryons at the drag epoch, when the baryons are released from the drag of the photons slightly before decoupling, leaves an imprint not only in the CMB anisotropies but also in the clustering of galaxies. By measuring the galaxy power spectrum together with CMB anisotropies, the value of $r_{\mathrm{d}} = 147.09 \pm 0.26 $ Mpc can be precisely determine \cite{Planck:2018vyg}. The value of this peak depends only on the initial conditions from the primordial universe, whereas the position of this peak relies on the angular diameter distance to the observed objects from the late universe. As such, BAO measurements can be used as a very standard ruler to measure distances. Preliminary data points from observations are typically represented in a dimensionless form as follows:
\begin{align}
    \frac{D_\mathrm{H}(z)}{r_{\mathrm{d}}} &= \frac{c}{H(z) r_{\mathrm{d}}}, \\
    \frac{D_\mathrm{M}(z)}{r_{\mathrm{d}}} &= \frac{D_\mathrm{L}(z)}{(1+z) r_{\mathrm{d}}} = \frac{(1+z) D_\mathrm{A}(z)}{r_{\mathrm{d}}}, \\
    \frac{D_\mathrm{V}(z)}{r_{\mathrm{d}}} &= \left[ z D_\mathrm{M}^2(z) D_\mathrm{H}(z) \right]^{1/3} / r_{\mathrm{d}},
\end{align}
where $D_\mathrm{H}(z)$, $D_\mathrm{A}(z)$, and $D_\mathrm{V}(z)$ represent the Hubble distance, angular diameter distance, and volume-averaged distance, respectively.

For the BAO data, we adopt the recent DESI BAO data from \cite{DESI:2024mwx} (the BAO data released by DESI in April 2024), along with some previous BAO data from SDSS and WiggleZ, which can be found in Table~III of \cite{Yang:2024kdo}. Moreover, we assume that the covariance matrix of all the data points is diagonal. In the following, the BAO dataset will specifically refer to the combination of DESI data and previous BAO data.

\textbf{Cosmic Chronometers (CC)}-- Through the measurement of age difference $\Delta t$ between two passively-evolving galaxies, which share the same formation time but are separated by a very small redshift difference $\Delta z$, it is possible to estimate the Hubble parameter $H(z)$ without assuming any specific cosmological model. Thus, the Hubble parameter can be directly extracted through the differential age method as 
\begin{equation}
    H(z)=-\frac{1}{1+z}\frac{\mathrm{d} z}{\mathrm{d} t}.
\end{equation}
For the CC data, we adopt the widely-used compilation of 31 data points, which can be found in Table~I of \cite{Mukherjee:2021ggf}.

\textbf{Redshift-space distortion (RSD)}-- Through the measurement of RSD, the distortions in the two-point correlation function caused by the Doppler effect of galaxy peculiar velocities associated with the gravitational growth of the LSS inhomogeneities, we can obtain information about the growth rate $f$ of the LSS. If we ignore the effect of massive neutrinos \cite{Villaescusa-Navarro:2017mfx}, the matter density contrast can be decomposed to the scale-independent part and scale-dependent part (initial condition), and  $\delta_{\mathrm{m}}(z,k) = D(z) \delta_{\mathrm{m}}(0,k)$, where $D(z)$ is the scale-independent linear growth function, while $\delta_{\mathrm{m}}(0,k)$ is the scale-dependent initial density contrast. However, since in this work we will consider specific modified gravity theories, it is essential to take their imprints on the matter perturbation into account. Typically, gravity modifications are reflected into an effective gravitational constant or a modified gravity parameter given by $\mu(k,a)=G_{\mathrm{eff}}/G$ \cite{Ishak:2018his,Koyama:2015vza}. In this case, the Poisson equation can be modified as
\begin{equation}
   k^2 \Psi = - 4 \pi G a^2 \mu(k,a) \Bar{\rho}_{\mathrm{m}} \delta_{\mathrm{m}}.
\end{equation}
Through the perturbed conservation equation and Poisson equation for the matter component in an expanding universe, we can express the modified evolution equation for the density contrast as
\begin{equation}
    \Ddot{\delta}_{\mathrm{m}}+2H\Dot{\delta}_{\mathrm{m}}-\frac{3}{2} \mu \Omega_{\mathrm{m}}H^2\delta_{\mathrm{m}}=0,
    \label{eq:matter_pert}
 \end{equation}
where $\Omega_{\mathrm{m}}(z)$ is the matter density parameter, with dots denoting time derivatives.

We define the scale-independent linear growth rate $f(z)$ of the LSS as
\begin{equation}
    f(z) \equiv -\frac{\partial \ln \delta_{\mathrm{m}}(z,k)}{\partial \ln (1+z)}=-\frac{\mathrm{d} \ln D(z)}{\mathrm{d} \ln (1+z)}.
\end{equation} By using the relation $\mathrm{d} z / \mathrm{d} t=-(1+z)H(z)$,   Eq.~\eqref{eq:matter_pert} can be rewritten as an evolution equation for the growth rate $f(z)$, namely 
\begin{equation}
\begin{aligned}
        -\frac{\mathrm{d} f}{\mathrm{d} \ln{(1+z)}}+f^2+\Big (-\frac{H'(z)(1+z)}{H}+2 \Big )f=\frac{3}{2}\mu \Omega_{\mathrm{m}},
\end{aligned}
\end{equation}
where primes represent the derivative with respect to the redshift $z$. Thus, we can reconstruct the modified gravity parameter $\mu$ through the relation
\begin{equation}
    \mu(z)=\frac{2}{3\Omega_{\mathrm{m}}}\Bigg \{-(1+z)f'+f^2+\Big (-\frac{H'(1+z)}{H}+2 \Big )f \Bigg \},
    \label{eq:mu_growth_rate}
\end{equation}
by using the growth rate data $f(z)$ as well as the background-level $H(z)$ data. In GR, this parameter should be $1$, thus, from the reconstructed deviation of $\mu(z)$ from $1$, we can test whether there are deviations from  GR. In the following, we will fix the evolution of $\Omega_{\mathrm{m}}(z)$. The RSD data we use in this paper can be found in Table 1 of \cite{Avila:2022xad}.

\subsection{Method}

The reconstruction method we apply in this work is known as Gaussian process regression \cite{Shafieloo:2012ht,Holsclaw:2010sk,Holsclaw:2010nb,Seikel:2012uu}. Gaussian process regression allows us to reconstruct the function and its derivatives in a model-independent manner from observational data points, utilizing a chosen kernel covariance function. In particular, consider that one desires to reconstruct $g(z)$, and has a set of observation data $(X,Y,\mathrm{COV}_Y)$ along with $(\mathrm{d}X,\mathrm{d}Y,\mathrm{COV}_{\mathrm{d}Y})$. Here, $X$ and $ Y $ represent the redshift and the values of the observational data $ g(z_i) $, respectively, while $ \mathrm{COV}_Y $ is a $ n_1 \times n_1 $ covariance matrix of $ g(z_i)$, where $ n_1 $ is the number of observations. Similarly, $ \mathrm{d}X $ and $ \mathrm{d}Y $ contain information on the observed redshift and the values of the first derivative data $ g'(z_i) $, $ \mathrm{COV}_{\mathrm{d}Y} $ is a $ n_2 \times n_2 $ covariance matrix of $ g'(z_i) $ with $ n_2 $ the number of the observed $ g'(z) $. In order to reconstruct $g(z)$ smoothly along with its derivatives $g'(z)$ and $g''(z)$, we follow the following steps:
\begin{enumerate}
    \item Acquire the observational data sets $(X, Y, \mathrm{COV}_Y)$ and $(\mathrm{d}X, \mathrm{d}Y, \mathrm{COV}_{\mathrm{d}Y})$ from cosmological observations.
    \item Select a specific kernel function.
    \item Use the Gaussian process to reconstruct the evolution of $ g(z)$ and its derivatives.
    \item Obtain the reconstructed functions $ g(z) $, $ g'(z)$, and $ g''(z) $, along with their covariance matrices.
\end{enumerate}
It should be noted that $(\mathrm{d}X, \mathrm{d}Y, \mathrm{COV}_{\mathrm{d}Y})$ is not mandatory, and Gaussian process regression can still be performed without it; however, the uncertainty in the reconstructed derivative will become larger.

For the kernel function, the squared exponential kernel covariance function is chosen, which is the most used kernel function in Gaussian process, defined as
\begin{equation}
    k\left(x, x^{\prime}\right)=\sigma_{f}^{2}\exp \Big [-\frac{\left(x-x^{\prime}\right)^{2}}{2 l^{2}} \Big ],
\end{equation}
which characterizes the relationship of the reconstructed function between two target points $x$ and $x^{\prime}$. The $\sigma_f$ and $l$ are the hyperparameters \cite{Seikel:2012uu}. During the reconstruction, these hyperparameters will be continuously optimized to achieve the best fit. Meanwhile, the squared exponential kernel covariance function is infinitely differentiable, making it particularly convenient and reliable for reconstructing the derivatives.

Specifically, in this paper, we want to reconstruct the Hubble parameter $H(z)$ and its derivative $H'(z)$. However, from the SNe data we can only obtain information about the comoving distance $D_\mathrm{M}(z)$. Fortunately, using the relation 
\begin{equation}
    \begin{aligned}
        H(z)&=\frac{c}{D'_\mathrm{M}(z)}, \\
        H'(z)&=-\frac{cD''_\mathrm{M}(z)}{(D'_\mathrm{M}(z))^2},
    \end{aligned}
\end{equation}
alongside  Gaussian process regression, through the reconstruction of the observed comoving distance $D_\mathrm{M}(z)$ and its first and second derivative,  we can acquire the Hubble parameter and its derivative. Meanwhile, the covariance matrix between the reconstructed $D'_\mathrm{M}(z)$ and $D''_\mathrm{M}(z)$ should also be transferred to the covariance matrix between $H(z)$ and $H'(z)$, through the relation
\begin{equation}
    \mathrm{COV}(H,H')=\mathcal{J} \cdot \mathrm{COV}(D'_\mathrm{M},D''_\mathrm{M}) \cdot \mathcal{J}^T,
\end{equation}
where $\mathcal{J}$ is the Jacobian matrix between $H$, $H'$ and $D'_\mathrm{M}$, $D''_\mathrm{M}$, and $\mathcal{J}^T$ is its transpose, defined as 
\begin{equation}
    \begin{aligned}
        \mathcal{J}&=\left( \begin{array}{cc}
           \frac{\partial H}{\partial D'_\mathrm{M}}  & \frac{\partial H}{\partial D''_\mathrm{M}}  \\
            \frac{\partial H'}{\partial D'_\mathrm{M}}  & \frac{\partial H'}{\partial D''_\mathrm{M}} 
        \end{array}
        \right ) 
        =\left( \begin{array}{cc}
           -\frac{c}{(D'_\mathrm{M})^2}  & 0  \\
            \frac{2cD''_\mathrm{M}}{(D'_\mathrm{M})^3}  & -\frac{c}{(D'_\mathrm{M})^2} 
        \end{array}
        \right ).
    \end{aligned}
\end{equation}
 
Additionally, we can directly reconstruct the Hubble parameter through the observed CC $H(z)$ data. If we want to combine the $D_\mathrm{M}(z)$ data with the $H(z)$ data, we can treat the $H(z)$ data as the data of the first derivative of $D_\mathrm{M}(z)$, and then apply the same procedure as what we did before. For RSD data, the $f(z)$ function can be directly reconstructed from the observations, but if we want to reconstruct $\mu$, we still need the background $H(z)$ data. Hence, when using the RSD data to reconstruct $\mu$ we will combine it with the reconstructed $H(z)$ function from the previous procedures.

After obtaining the reconstructed $H(z)$ and $H'(z)$, through the background Friedmann equation we can obtain the dark energy EoS parameter
\begin{equation}
   \ w=\frac{2H'(1+z)H-3H^2}{3H^2(1-\Omega_{\mathrm{m}})}.
    \label{eq:w_background}
\end{equation}
Furthermore, we  can  define the normalized effective dark energy density parameter $X(z)$ as
\begin{equation}
    X(z)=\frac{\rho_{de}(z)}{\rho_{de}(0)}=\frac{3H^2(1-\Omega_{\mathrm{m}})}{3H_0^2(1-\Omega_{\mathrm{m}0})},
    \label{eq:X(z)_define}
\end{equation}
where $\rho_{de}$ may include the information about modifications of gravity, scalar fields, etc. The value of the current matter density parameter measured by Planck \cite{Planck:2018vyg} is $\Omega_{\mathrm{m}0}=0.3153 \pm 0.0073$. Note that if the value of $X(z)$ becomes smaller than $0$  during the evolution, implies that the effective dark energy density shall become negative. 

We also comment that in the above calculations, especially for the matter perturbations, we have used the assumption of conservation equations for matter component, which implies that the interaction between dark energy and matter is no longer under consideration, leading to
\begin{equation}
    \Omega_{\mathrm{m}}(z)=\Omega_{\mathrm{m}0}H_0^2(1+z)^3/H^2(z).
    \label{eq:om_no_interaction}
\end{equation}
This relation can be used for the reconstruction of $w$, $X$ and $\mu$ through Eq.~\eqref{eq:w_background}, \eqref{eq:X(z)_define} and \eqref{eq:mu_growth_rate}.

\subsection{Results and Discussion}

\begin{figure*}[htbp]
    \centering
	\includegraphics[width=0.95\textwidth]{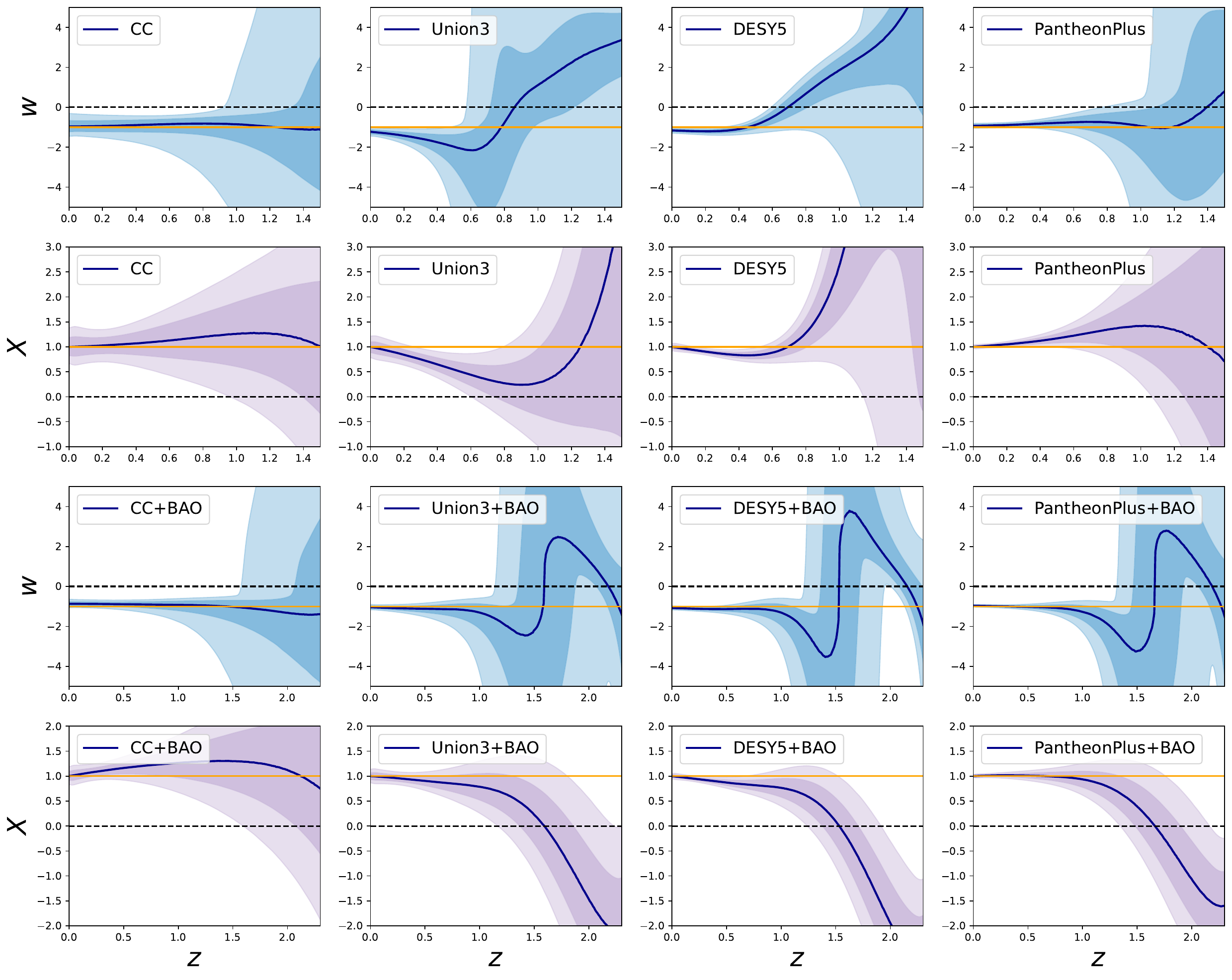}
	\caption{{\it{The mean value of the reconstructed  dark energy EoS parameter $w(z)$ and the normalized dark energy density parameter $X(z)$, along with the $1\sigma$ and $2\sigma$ uncertainties, for CC/SNe only and CC/SNe + BAO datasets. For comparison, we have added the yellow solid line, which shows the value of the parameters predicted by the $\Lambda$CDM paradigm. Finally,  the black dashed line is added for convenience, and marks  whether $w$ and $X$   change sign.}}}
	\label{fig:1}
\end{figure*}

Let us now present the results. In Fig.~\ref{fig:1} we show the reconstructed $w(z)$ and $X(z)$ for CC/SNe-only and CC/SNe + BAO data sets. Due to the limitation of the observational data at high redshift for the SNe dataset, we only plot the evolution of the parameters within the redshift range ($0$, $1.5$). We find that for the reconstructed results of the CC-only case, the $\Lambda$CDM model lies well within the 1$\sigma$ allowed range, with only a slight preference for the quintom-B behavior. 
The combination of BAO and CC datasets shows that these two observations are in mutual agreement.
However, the above consistency is not evident between SNe and BAO datasets, and even different SNe observations give slightly different results.

\begin{figure*}[htbp]
    \centering
	\includegraphics[width=\textwidth]{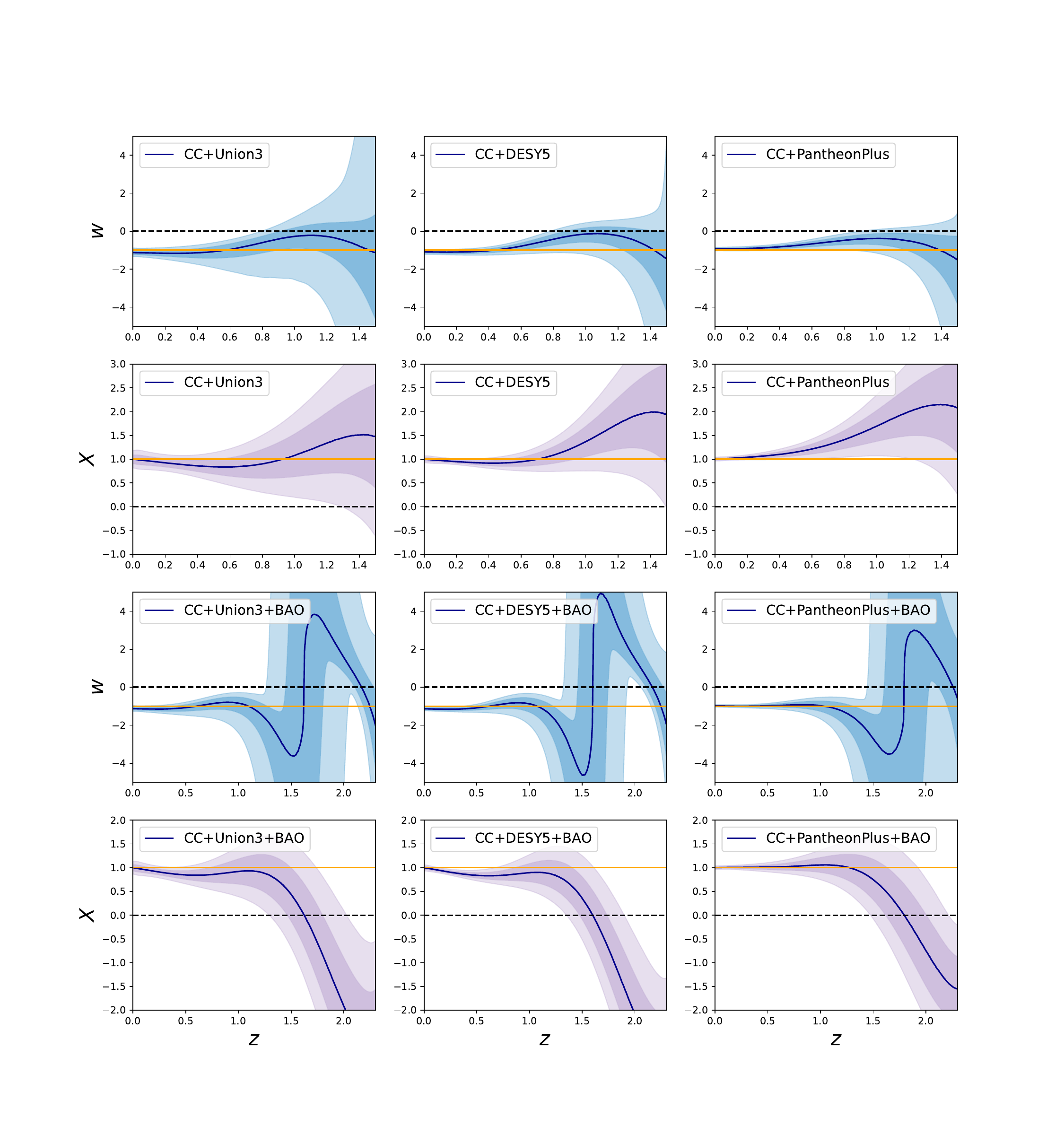}
	\caption{{\it{The mean value of the reconstructed  dark energy EoS parameter $w(z)$ and the normalized dark energy density parameter $X(z)$, along with the $1\sigma$ and $2\sigma$ uncertainties for CC + SNe and CC + SNe + BAO datasets. For comparison, we have added the yellow solid line, which shows the value of the parameters predicted by the $\Lambda$CDM paradigm. Finally,  the black dashed line is added for convenience, and marks  whether $w$ and $X$   change sign.}}}
	\label{fig:2}
\end{figure*}

\begin{figure*}[htbp]
    \centering
	\includegraphics[width=\textwidth]{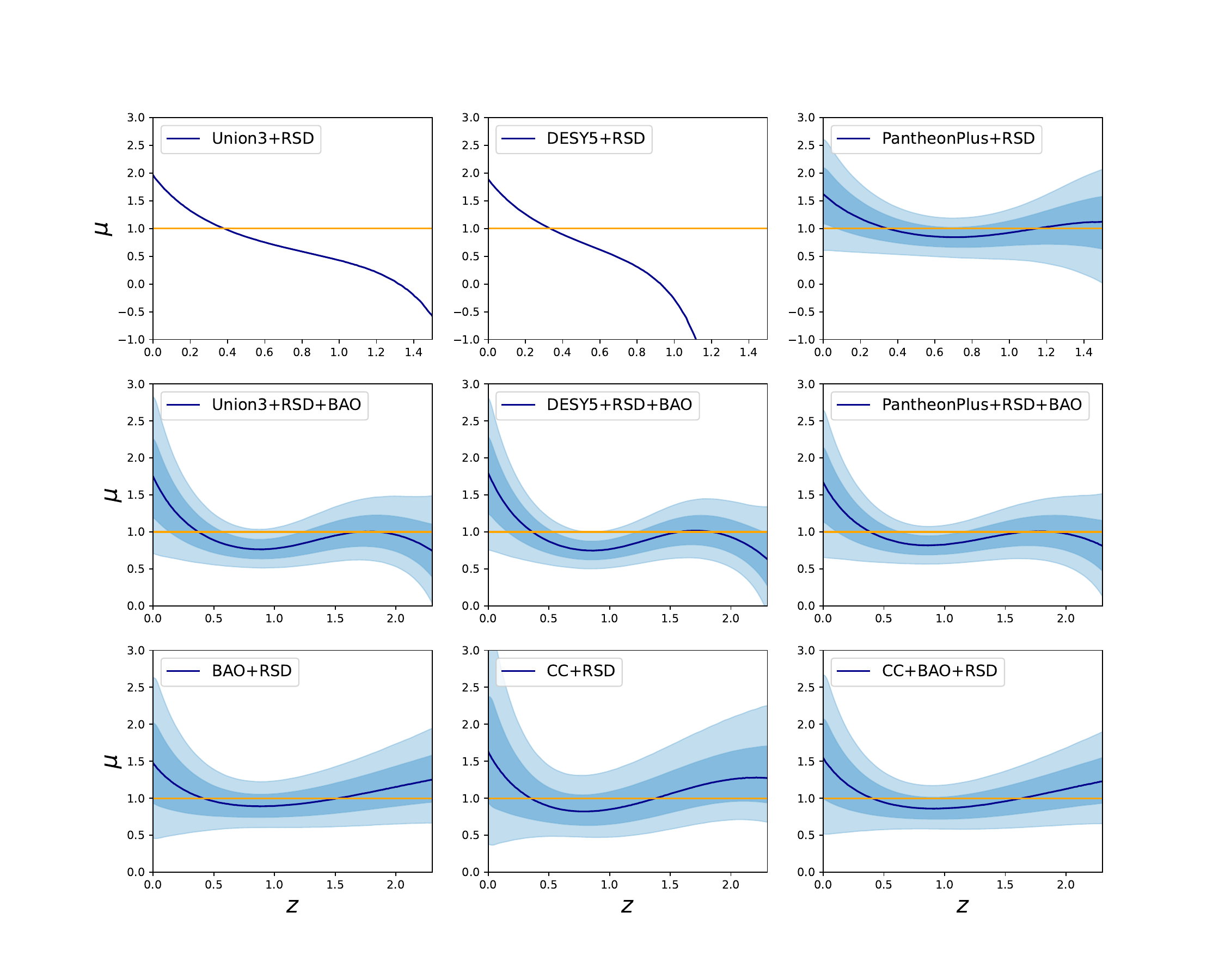}
	\caption{{\it{The mean value of the reconstructed modified gravity parameter $\mu(z)$ along with the $1\sigma$ and $2\sigma$ uncertainties for CC/SNe + RSD and CC/SNe + RSD + BAO datasets. For comparison, we have added the yellow solid line, which shows the value of the parameters predicted by the $\Lambda$CDM paradigm. Finally,  the black dashed line is added for convenience, and marks  whether $w$ and $X$   change sign.}}}
	\label{fig:3}
\end{figure*}

The results for the SNe-only dataset show that the dark energy EoS parameter $w$ smoothly crosses zero. Note that the results of $X$ for SNe indicate that there is no change of sign for the dark energy density in the course of evolution. Hence the smooth sign change of $w$ can only be attributed to the effective dark energy pressure shifting from positive to negative as the universe expands. Additionally, $w$ exhibits quintessence behavior for the PantheonPlus dataset, whereas it shows phantom behavior for Union3 and DESY5 datasets, at redshifts lower than $0.4$. Note that $\Lambda$CDM still falls very precisely within 1$\sigma$ confidence level for PantheonPlus, but no longer for Union3 or DESY5. Even for the Union3 data, at redshift of about 0.3, the $\Lambda$CDM model is already beyond the 2$\sigma$ range. Meanwhile, once the BAO dataset is combined with the SNe data, the results become quite different. For all combined datasets, we find that $w$ crosses zero due to the sign change of the effective dark energy density $X$ during the expansion, and similar results were also  found in \cite{DESI:2024aqx}. Specifically, for Union3 + BAO and DESY5 + BAO, $w$ displays phantom behavior in the later universe, while the PantheonPlus + BAO exhibits quintom-like behavior. Through the reconstruction of $X$, we find that in the range of redshift greater than $1.5$, the $\Lambda$CDM scenario has exceeded the 2$\sigma$ range allowed by the reconstruction results, and especially DESY5 + BAO datasets lead to the most significant deviation for the $\Lambda$CDM model.

In Fig.~\ref{fig:2} we show the reconstructed $w(z)$ and $X(z)$ for CC + SNe and CC + SNe + BAO datasets. After adding SNe data to the CC data, we find that $w$ exhibits quintom-A behavior for CC + Union3/DESY5, while for CC + PantheonPlus  datasets it shows quintessence behavior, for redshifts smaller than $1$. On the other hand, when the redshift is larger than $1.2$, a quintom-A behavior appears, and in this case $\Lambda$CDM no longer falls within the 1 $\sigma$ region. Nevertheless, if we add BAO data further  we   find different behavior. For CC + Union3/DESY5 + BAO datasets in the range of redshifts less than $1.5$, we find that $w$ crosses $-1$ twice as the universe evolves, the first time as quintom-B and the second time as quintom-A, and finally both cases present signatures of phantom dark energy at current time. Similar deviation from the $\Lambda$CDM model is also observed in the reconstructed $X$, as we have discussed above. We can thus conclude that it is the existence of BAO data that leads to the energy density of dark energy showing a trend of sign-changing.

We proceed by considering the RSD data in our analysis. The results of $\mu(z)$, obtained using \eqref{eq:mu_growth_rate}, are displayed in Fig.~\ref{fig:3}. Interestingly, we find a consistent behavior of $\mu$ across almost all results, and we find no strong evidence that GR needs to be significantly modified, only small corrections are allowed. Although for the reconstruction results of Union3 and DESY5  we find that there are   time intervals where $\Lambda$CDM (or GR) has exceeded the confidence interval of 2$\sigma$, in most cases $\Lambda$CDM scenario  can still fall well within the confidence interval of 2$\sigma$. Meanwhile, for the value of $\mu$ at the current time, all reconstructions seem to tend to be slightly greater than $1$.

At low redshifts ($z<0.6$), although there are cases where the $\Lambda$CDM model does not fall well within the 1$\sigma$ range, it is almost always within the 2$\sigma$ range, except for the Union3-only result where a signal beyond 2$\sigma$ can be observed for a particular redshift range. However, after a joint analysis of the SNe and BAO data, we found a signal that significantly exceeds the $\Lambda$CDM paradigm at high redshifts ($z>1.5$), especially the combination of DESY5 and BAO. Additionally, all the reconstruction results of $X$ show a common decreasing trend in the range of redshift greater than 1.5, thus there is a high probability that it will cross $0$. We should emphasize that since the information we acquire about $H$ and $H'$ is actually derived from the first and second derivatives of $D_\mathrm{M}$, the uncertainty of $w$ or $X$ will become relatively larger.

We mention that the signal beyond the $\Lambda$CDM scenario at high redshifts within this increased error is a robust signal for dynamical dark energy. Therefore, we conclude that there is no strong evidence for the presence of dynamical dark energy at low redshift, but there is a strong tendency towards dynamical dark energy at high redshift. The inclusion of BAO data appears to induce a sign change in the effective dark energy density, thereby leading to a distinctly different behavior of $w$ compared to results without BAO data at high redshift. Meanwhile, the results from different SNe datasets reveal partial consistency, as do the CC and BAO datasets. Finally, the inclusion of perturbation-level RSD data indicates that one needs only    minor corrections to the standard gravity theory, which is a very strong constraint for judging the feasibility of different modified gravity models.

Based on the above results, we summarize the behaviors obtained from different datasets:
\begin{itemize}
\item \textbf{BAO}: The reconstruction results are consistent with those of CC, and tend to exhibit the effectively negative dark energy density in the high-redshift phase.
\item \textbf{SNe}: The results of SNe alone tend to positive energy density, except the PantheonPlus dataset, and the EoS parameter directly shows an upward trend in the redshift phase of 0.8-1.5. When combined with the BAO dataset, the results reveal an oscillation tendency, and the EoS parameter exceeds -1 more than once.
\end{itemize}

Below, we divide the evolution characteristics of $w$ or $X$ in the reconstruction result into the following categories:
\begin{enumerate}
    \item From most of the reconstruction results of $X$, we find that the effective dark energy density has a common tendency to become negative at high redshifts. Hence, we refer to it as the negative-energy dark energy behavior.
    \item For most of the results, we also find that the dark energy EoS parameter crosses the cosmological boundary $-1$ more than once, and this belongs to the broad class of oscillating dark energy  \cite{Feng:2004ff, Pan:2017zoh}.
    \item Furthermore, we notice that for some cases, the effective dark energy pressure is positive in the early universe, and then it becomes negative to dominate the late-time cosmic acceleration. Therefore,  we call it late-dominated dark energy behavior.
\end{enumerate}

\section{Dynamical interpretation} \label{theory}

In order to describe the behavior of dynamical dark energy we need to add   additional degree(s) of freedom compared to the standard $\Lambda$CDM scenario. The dynamical evolution of dark energy can be naturally realized in models involving multiple scalar fields, which arise in particle physics or string theory. In the literature, a wide variety of dark energy constructions have been proposed, including quintessence, phantom, K-essence, quintom, modified gravity, etc. In the following, we display    some specific   scenarios that can describe the  dark energy sector.

\subsection{Negative-energy dark energy}

Firstly, for negative-energy dark energy behavior    will lead   the matter density parameter to  become larger than $1$. Note that in the definition of the matter density parameter $\Omega_{\mathrm{m}}=\rho_{\mathrm{m}}/(\rho_{\mathrm{m}}+\rho_{de})$, one usually thinks of dark energy as a real component of our universe, whose density is necessarily non-negative, but in fact it may arise  due to pure gravity effects or due to the interaction between gravity and matter. Thus, what we define as the energy density of dark energy is not necessarily the energy density of an actual component, but it can have an effective nature and hence it may be negative. If we consider that the accelerated expansion of the late universe is driven by a minimally coupled scalar field, then its energy density must be non-negative. On the other hand, if we try to explain the negative dark energy density in terms of the scalar field model, then we must introduce the non-minimal coupling of the scalar field with gravity, in which case the dark energy density includes the effect of gravity. However, it is more natural to describe this phenomenon in the framework of modified gravity theories. This can be realized by introducing additional gravitational terms, thus avoiding the need to introduce other extra components or couplings. These additional terms exhibit effective properties, leading to the accelerated expansion of the universe.

$\mathbf{f(R)}$ \textbf{gravity} theory is a typical and well-established modified gravity theory. In this framework, the Einstein-Hilbert action is generalized by replacing the Ricci scalar $R$  with a generic function $f(R)$ \cite{Starobinsky:1980te, Capozziello:2002rd, DeFelice:2010aj}:
\begin{align}
    S = \int \mathrm{d}^4 x \sqrt{-g} \left[ \frac{1}{2} f(R)  \right] ~.
\end{align}

In the cosmological background, the extra gravitational term in the   field equations relative to GR can be transferred to the matter part as an effective matter component. These effective terms introduce corresponding effective energy density and effective pressure of the form
\begin{equation}
\begin{aligned}
 \rho_{f(R)}=&\frac{1}{f_R}\left[\frac{1}{2}(f-Rf_R)-3H\dot{R}f_{RR} \right] ~, \\
 p_{f(R)}=&\frac{1}{f_R}(2H\dot{R}f_{RR}+\Ddot{R}f_{RR})+\frac{1}{f_R}\left[\dot{R}^2f_{RRR} -\frac{1}{2}(f-R f_R)\right] ~,
\end{aligned}
\end{equation}
with $R=12H^2+6\dot{H}$ and $f_R={\rm d}f/{\rm d}R$, $f_{RR}={\rm d}^2f/{\rm d}R^2$. Hence,  the effective equation-of-state parameter is given by
$w_{f(R)}=p_{f(R)} /  \rho_{f(R)}$. The flexibility in choosing the functional forms of $f(R)$ enables a wide range of phenomenological models, which can be constrained by observational data.

$\mathbf{f(T)}$ \textbf{gravity}, similarly, a modified theory of gravity can be obtained by starting from other geometric theories equivalent to GR. $f(T)$ gravity is based on the torsion scalar $T$ and the  teleparallel equivalent of general relativity (TEGR) formulation \cite{Cai:2015emx, Bahamonde:2021gfp, Krssak:2018ywd}. Using the tetrad field $h^{a}{}_{\mu}$, the general action of $f(T)$ gravity is:
\begin{equation}
 S = \int \mathrm{d}^{4} x h \left[ \frac{1}{2} f(T)  \right] ~,
\end{equation}
where $h=\det\left(h^{a}{}_{\mu}\right)=\sqrt{-g}$. Similar to $f(R)$ gravity, one can also introduce the effective energy density and  pressure under $f(T)$ gravity as
\begin{equation}
\begin{aligned}
 \rho_{f(T)}=&Tf_T-\frac{1}{2}(f+T) ~, \\
 p_{f(T)}=&\frac{f-Tf_T+2T^2f_{TT}}{2f_T+4Tf_{TT}}  ~,
\end{aligned}
\end{equation}
with $f_T={\rm d}f/{\rm d}T$, $f_{TT}={\rm d}^2f/{\rm d}T^2$. Thus, 
the effective EoS of dark energy   is given by
\begin{equation}
\label{wofg}
w_{f(T)} \equiv \frac{p_{f(T)}}{\rho_{f(T)}}=\frac{f-Tf_{T}+2 T^2 f_{T T}}{\left[f_{T}+2 T f_{T T}\right]\left[2T f_{T}-f-T\right]}. 
\end{equation}

$\mathbf{f(Q)}$ \textbf{gravity} is based on the concept of non-metricity, where the gravitational interaction is described by the non-metricity scalar $Q$. Under the coincident gauge, the Friedmann equations of $f(Q)$ gravity have the same form as $f(T)$ gravity \cite{BeltranJimenez:2019tme,Anagnostopoulos:2021ydo}. Nevertheless, there are different connection branches in the $f(Q)$ theory, leading to different cosmic dynamics. In the most general case  we have
\begin{equation}
    Q=3\left(-2H^2+3C_3H+\frac{C_2}{a^2}H-(C_1+C_3)\frac{C_2}{a^2}+(C_1-C_3)C_3\right)
\end{equation}
where $C_1$, $C_2$, $C_3$ are   functions of time, which can be parametrized   using a free temporal function $\gamma$ \cite{Yang:2024tkw}. For the convenience of representation, we generally introduce $f(Q)=Q+F(Q)$ with $F_Q={\rm d}F/{\rm d}Q$, $F_{QQ}={\rm d}^2F/{\rm d}Q^2$. Similarly, the effective dark energy  density and pressure are respectively given by
\begin{align}
\rho_{f(Q)}=&-\frac{1}{2}F+(\frac{1}{2}Q-3H^2)F_Q-\frac{3}{2}\dot{Q}(C_3-\frac{C_2 }{a^2})F_{QQ},  \\    
p_{f(Q)}=&\frac{1}{2}F+(2\dot{H}+3H^2-\frac{1}{2}Q)F_Q -\frac{1}{2}\dot{Q} 
(-4H+3C_3+\frac{C_2}{a^2})F_{QQ}, 
\end{align}
while the EOS parameter is $w_{f(Q)}=p_{f(Q)}/ \rho_{f(Q)}$.

It is worth noting that in the non-coincident gauge scenario, the matter conservation equation    deviates from the standard form. This result arises from an effective interaction between the connection structure of the geometry and the matter sector, similar to the case of interacting dark energy. 

These theories offer alternative mechanisms for realizing negative-energy dark energy. In particular, in order to achieve a sign-changing dark energy density, one must have the crossing conditions at the crossing redshift $z_\mathrm{c}$, namely
\begin{equation}
    \rho_{\mathrm{de}(z_\mathrm{c})}=0 \quad \quad \text{and}  \quad \quad \frac{\mathrm{d} \rho_{\mathrm{de}}}{\mathrm{d} z} \Big | _{z=z_\mathrm{c}}<0.
\end{equation}
Let us use $f(T)$ gravity as an example to describe the negative-energy dark energy. Note that we only need to make minor modifications to general relativity, which implies that a reasonable modification  can always be effectively described in the   form $f(T)=\alpha T+bT^2$, with $b$  a small parameter. In this specific model, the crossing conditions can be rewritten as
\begin{equation}
    \begin{aligned}
    \rho_{f(T)}(z_{\mathrm{c}})=&T_{\mathrm{c}}(\alpha+2bT_{\mathrm{c}})-\frac{1}{2}\Big( (\alpha+1)T_{\mathrm{c}}+bT_{\mathrm{c}}^2 \Big) \\
    =&\Big(\frac{\alpha-1}{2}+\frac{3b}{2}T_{\mathrm{c}} \Big)T_{\mathrm{c}}=0,  \\ 
    \frac{\mathrm{d} \rho_{f(T)}}{\mathrm{d} z} \Big | _{z=z_{\mathrm{c}}}=&\Big (\frac{\alpha-1}{2}+3bT_{\mathrm{c}} \Big)\frac{\mathrm{d}T}{\mathrm{d}z} \Big | _{z=z_{\mathrm{c}}}\\
    =&\frac{3b}{2}T_{\mathrm{c}} \frac{\mathrm{d}T}{\mathrm{d}z} \Big | _{z=z_{\mathrm{c}}} <0,
    \end{aligned}
\end{equation}
where in the second condition  we have used the relation from the first condition, and $T_{\mathrm{c}}$ represents the value of $T$ at the crossing redshift, thus $T_{\mathrm{c}}=T(z=z_{\mathrm{c}})$. Since we always have $T=-6H^2<0$, the crossing conditions can be written as
\begin{equation}
    \begin{aligned}
    \alpha-1+3bT_{\mathrm{c}}&=0  \\
    b\frac{\mathrm{d} T}{\mathrm{d} z}\Big | _{z=z_{\mathrm{c}}} &>0. 
    \end{aligned}
    \label{eq:negative condition}
\end{equation}
From this relation  we can easily see that for $\alpha=1$  it is not possible to find a suitable value of $b$ that can satisfy both these conditions. Thus, to achieve negative dark energy density we require $\alpha\neq 1$. Meanwhile, note that we always have $\mathrm{d} T/\mathrm{d}z=\mathrm{d}(-6H^2)/\mathrm{d}z<0$ and $T<0$, hence in order to acquire a solution for Eq.~\eqref{eq:negative condition}, $b<0$ and $\alpha<1$ are required.

However, in order to obtain a healthy theory, it is not enough  to consider only the background evolution, but we also need to require that the perturbations of the theory be stable. We focus on the scalar perturbations and we write the perturbed metric in Newtonian gauge, namely
\begin{equation}
    d s^2=-(1+2\Psi)d t^2+a^2(1-2\Phi)\delta_{ij}dx^i dx^j.
\end{equation}
Since we are concerned with the stability of the theory, in the following we will ignore the other components, and we only concern the pure gravitational sector,  since if the gravitational sector itself is unstable  the matter content cannot cure the instability. For simplicity  we impose the zero-anisotropic-stress assumption, which implies that $\Psi=\Phi$. As it is shown in \cite{Chen:2010va},   the requirement of no anisotropic stress imposes another constraint on $f(T)$ models, namely that $f_{TT}(T) \simeq 0$. In our chosen functional form  this implies that $b \simeq 0$. The mode-expansion of $\Phi$ can be expressed in Fourier space as
\begin{equation}
    \Phi(t,\mathbf{x})=\int \frac{\mathrm{d}k^3}{(2\pi)^{2/3}} \tilde{\Phi}_k(t) e^{i\mathbf{k} \cdot \mathbf{x}}.
\end{equation}
Thus, considering the pure gravitational effect, the equation of motion for the scalar perturbation $\tilde{\Phi}_k$ can be expressed as \cite{Chen:2010va}
\begin{equation}
    \Ddot{\tilde{\Phi}}_k+\Gamma \dot{\tilde{\Phi}}_k +
    \omega^2 \tilde{\Phi}_k=0,
\end{equation}
where $\Gamma$ and $\omega^2$ depend on the background evolutions. Therefore, a model in which $\omega^2$ is negative will  be unstable. Note that in such a pure gravitational case  the field equations in $f(T)$ gravity   lead to a constant Hubble parameter $H$ and $T=-6H^2$, thus the expression of $\omega^2$ can be given as
\begin{equation}
    \omega^2=\frac{\frac{3H^2}{2}-\frac{f-T}{4}-36H^4 f_{TT}}{f_T-12H^2 f_{TT}},
\end{equation}
and    this   leads to
\begin{equation}
    \omega^2=\frac{\alpha-54H^2b}{\alpha-36H^2b}\cdot \frac{3}{2}H^2.
\end{equation}
As we observe, in order to acquire a healthy theory  we must have $\alpha>54H^2b$ or $\alpha<0$. As we mentioned before, $b \simeq 0$, and hence   the above expression can be simplified 
to $w^2=\frac{3}{2}H^2$, which is  always   non-negative. Thus, we conclude that the model we used will always be stable under the condition $b \simeq 0$. Furthermore, note that in $f(T)$ gravity  the effective gravitational constant is $G_{\mathrm{eff}} \approx G/f_T$ \cite{Zheng:2010am}, and therefore in our specific model  the modified gravity parameter $\mu$ can be expressed as
\begin{equation}
    \mu(z) \approx \frac{1}{f_T}=\frac{1}{\alpha+2bT(z)}=\frac{1}{\alpha-12bH^2(z)}.
\end{equation}

\begin{figure}[htbp]
    \centering
	\includegraphics[width=\textwidth]{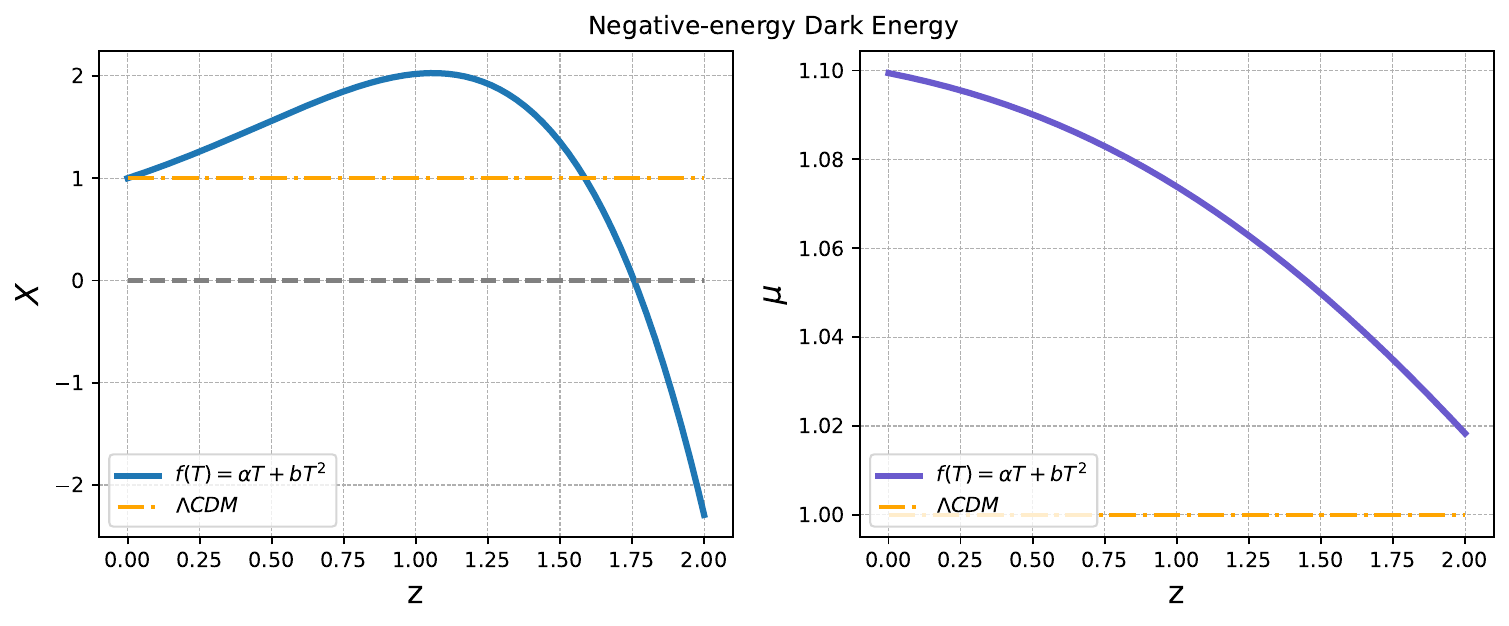}
	\caption{{\it{An example of negative-energy dark energy model within $f(T)$ gravity. The parameter values have been chosen as $\alpha=0.9$, $b=-1.5\times 10^{-7}$, where $\alpha$ is a dimensionless parameter, while $b$ has   dimensions of $H_0^{-2}$.}}}
	\label{fig:fT ne-de example}
\end{figure}

As an  example, let us  choose $\alpha=0.9$, $b=-1.5 \times 10^{-7}$. In this case,  the evolution of the normalized dark energy density $X$ and the modified gravity parameter $\mu$ are shown in Fig.~\ref{fig:fT ne-de example}. As we can see, with   suitable parameter choices we can achieve a negative-energy dark energy model within the formalism of modified gravity.

\subsection{Late-dominated dark energy}

For late-dominated dark energy behavior  we notice that at  early times the effective pressure is positive, acting in the same way as the composition of matter, causing the expansion of the universe to slow down. At later times the effective pressure becomes negative, acting as  dark energy, driving the acceleration of the expansion of the universe. Recall that for a single scalar field  its dark energy EoS parameter satisfies $-1 \leq w \leq 1$, which allows for $w$ to   cross $0$. Therefore, we will adopt a single scalar field to describe this dark energy model.

The \textbf{quintessence} scenario was first proposed by \cite{Ratra:1987rm,Wetterich:1987fm}, and it is described by an ordinary scalar field $\phi$ minimally coupled to gravity. The
action of quintessence dark energy is written as
\begin{equation}
    S = \int \mathrm{d}^{4}x \sqrt{-g} \left[-\frac{1}{2}(\nabla \phi)^{2}-V(\phi)\right],
\end{equation}
where $(\nabla \phi)^{2}=g^{\mu\nu}\partial_{\mu}\phi\partial_{\nu}\phi$ and $V(\phi)$ is the potential of the scalar field. In a flat FRW spacetime the energy density and pressure   of the scalar field is 
\begin{equation}
    \rho_\mathrm{Q} = \frac{1}{2}\dot{\phi}^{2}+V(\phi),
    p_\mathrm{Q} = \frac{1}{2}\dot{\phi}^{2}-V(\phi),
\end{equation}
and the evolution of the universe is described by
\begin{align}   
    H^{2} &= \frac{8\pi G}{3}\left[\frac{1}{2}\dot{\phi}^{2}+V(\phi)\right],\\
    \frac{\Ddot{a}}{a} &=  -\frac{8\pi G}{3}\left[\dot{\phi}^{2}-V(\phi)\right],
\end{align}
 and thus  the universe accelerates when
$\dot{\phi}^{2}\leq V(\phi)$. 
Then the EoS for the dark energy is given by
\begin{equation}
   w_\mathrm{Q} = \frac{p_\mathrm{Q}}{\rho_\mathrm{Q}}=\frac{\dot{\phi}^{2}-2V(\phi)}{\dot{\phi}^{2}+2V(\phi)},
\end{equation}
and as it can be seen it remains in the region $-1\leq w_\mathrm{Q}\leq 1$.
An exponential potential is typically used to accommodate the epoch  of current acceleration, and moreover it can possess cosmological scaling solutions in which the field energy density ($\rho_\mathrm{Q}$) is proportional to the fluid energy density ($\rho_{\mathrm{m}}$) \cite{Copeland:1997et,Barreiro:1999zs}, however    power-law   potentials can used too \cite{Caldwell:1997ii}.
Based on current observations, the energy density in the quintessence field should be consistent with the current critical energy density. This leads to the conclusion that the field
value at present is of the order of the Planck mass ($\phi_{0}~M_{\mathrm{P}}$), which is typical in most   quintessence models.

The \textbf{K-essence} scenario was inspired by k-inflation  to explain early universe inflation at high energies \cite{Garriga:1999vw}, in which the accelerated expansion arises  from modifications to the kinetic energy of the scalar field. Then \cite{Armendariz-Picon:2000nqq} applied it to dark energy, in order to explain the late-time acceleration. K-essence is characterized by a scalar field with a non-canonical kinetic energy.
In the most general case, the scalar-field action is a function of $\phi$ and $X \equiv -(1/2)(\nabla\phi)^{2}$ is
given by
\begin{equation}
    S = \int \mathrm{d}^{4}x \sqrt{-g} P(\phi, X).
    \label{eq:general single scalar field}
\end{equation}
This action   includes
quintessence scenario, where the Lagrangian density $P(\phi, X)$ corresponds to a pressure density. K-essence scenarios are generally restricted to the Lagrangian density of the form
\begin{equation}
    P(\phi, X) = f(\phi)\hat{P}(X).
\end{equation}
After an appropriate field definition \cite{Chiba:1999ka} one
chooses $\hat{P}(X)=-X+X^{2}$, and thus 
the energy density of the field $\phi$ is given by
\begin{equation}
    \rho = 2X\frac{\partial P}{\partial X}-P = f(\phi)(-X+X^{2}),
 \end{equation}   
while the EoS of the scalar field is given by
 \begin{equation}
    w_\mathrm{K} = \frac{p}{\rho}=\frac{1-X}{1-3X}.
 \end{equation}   
In this case, whether $w_\mathrm{K}$ varies or not depends on $X$, and in particular for $X = 1/2$  we obtain the EoS of a cosmological constant, i.e. $w_\mathrm{K} = -1$.
The EoS that gives rise to   accelerated expansion is $w_\mathrm{K}<-1/3$,
which translates into the condition $(X< 2/3)$.
 Furthermore, in \cite{Armendariz-Picon:2000nqq,Armendariz-Picon:2000ulo} 
 the authors extended the analysis
to more general forms of $\hat{P}(X)$, in order to solve the
coincident problem of dark energy.
Various aspects of K-essence and its applications to dynamical dark energy can be found in \cite{Scherrer:2004au,Creminelli:2008wc,Bonvin:2006vc,Myrzakulov:2012axz}.

\begin{figure}[htbp]
    \centering
	\includegraphics[width=0.7\textwidth]{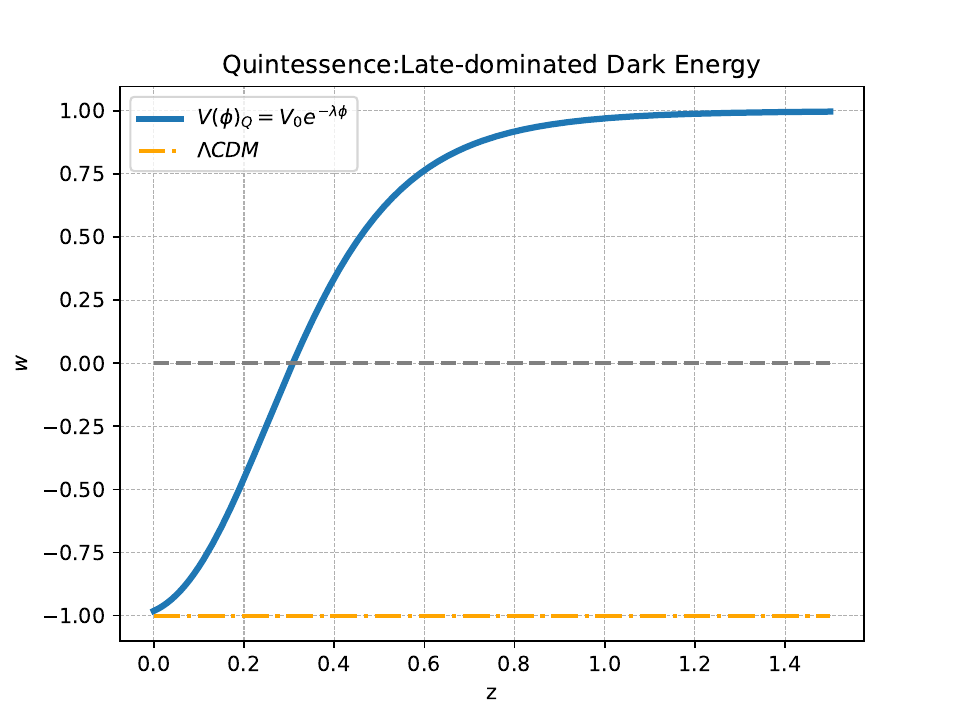}
	\caption{{\it{An example for late-dominated dark energy realization within quintessence scenario. The parameter values have been chosen as $V_0=1$ and $\lambda=\sqrt{3}$, where $V_0$ has  dimensions of  $H_0^2$ and $\lambda$ is a dimensionless parameter. We have set the initial conditions as $\phi(z=0)=-5.4$ and $d \phi/d z(z=0)=0.2$.}}}
	\label{fig:quintessence wde example}
\end{figure}

Let us use the simplest quintessence model as an example. We consider the exponential potential 
\begin{equation}
    V(\phi)=V_0 e^{-\lambda \phi},
\end{equation}
where $V_0,\lambda>0$. The stability of this model has been studied extensively in \cite{Andriot:2024jsh} and references therein.
Note that we can always introduce a constant field value $\phi_*$, satisfying $e^{\lambda \phi_*}=V_0$, so that $V(\phi)=e^{-\lambda (\phi-\phi_*)}$. Then, through the redefinition of the scalar field $\tilde{\phi}=\phi-\phi_*$, one can absorb the constant $V_0$ and  the potential is   only characterized by one free parameter, namely $\lambda$.
Without loss of generality  we choose $V_0=1$, and in order to acquire a suitable solution  we set $\lambda= \sqrt{3}$. Under such parameter selection, the evolution of the dark energy EoS parameter is shown in Fig.~\ref{fig:quintessence wde example}.

\subsection{Oscillating dark energy}

In this subsection  we propose a parametrization to describe   the oscillating dark energy evolution. However, in the case where the dark energy EoS parameter crosses $-1$  we must mention the No-Go theorem of dynamical dark energy,
which forbids the EoS parameter of a single perfect fluid or a single scalar field to cross the $-1$ boundary.
For example, consider a barotropic perfect fluid  which is described by pressure $p$, energy density $\rho$, and thus by EoS parameter $w=p/\rho$. Then the adiabatic sound speed is determined by 
\begin{equation}
    c_{\mathrm{s}}^2=\frac{p'}{\rho '}=w-\frac{w'}{3\mathcal{H}(1+w)},
\end{equation}
where the prime denotes the derivative with respect to the conformal time $\eta$,
with $ad\eta=dt$, and $\mathcal{H}$ is the conformal Hubble parameter.
Note that the sound speed of a single perfect fluid is apparently divergent when $w$ crosses -1, which leads to instability in DE perturbations. Similarly, one can also prove that the sound speed for a non-barotropic fluid will also become divergent when crossing the cosmological constant boundary, and finally this is also true   for a generic single scalar fluid  whose action is defined by the 
K-essence form Eq.~\eqref{eq:general single scalar field}. 
Additionally, one can also find that the dispersion relation becomes divergent at the crossing point as well. A detailed proof of the No-Go theorem can be found in \cite{Cai:2009zp}.

In summary, to construct a model capable of producing a quintom scenario, it is necessary to break certain constraints imposed by the No-Go theorem. For instance, one approach involves utilizing two scalar fields, where one acts as quintessence and the other as a phantom field. Individually, the EoS for each component does not need to cross the cosmological constant boundary, thereby ensuring the stability of their classical perturbations. Nevertheless, the combined dynamics of these two components can result in a quintom behavior \cite{Feng:2004ad}.

The action of \textbf{quintom} dark energy, combining a quintessence field $\phi$ with a phantom field $\sigma$, can be described by \cite{Feng:2004ad}
\begin{equation}
    S = \int \mathrm{d}^{4}x \sqrt{-g} \left[ -\frac{1}{2}(\nabla \phi)^{2}  +\frac{1}{2}(\nabla \sigma)^{2} -V(\phi, \sigma)\right].
\end{equation}
Thus, the effective energy density $\rho_\mathrm{Q}$ and the effective pressure $p_\mathrm{Q}$    are given by 
\begin{align}
    \rho_\mathrm{Q}& = \frac{1}{2}\dot{\phi}^{2}-\frac{1}{2}\dot{\sigma}^{2}+ V(\phi, \sigma), \\
    p_\mathrm{Q}& = \frac{1}{2}\dot{\phi}^{2}-\frac{1}{2}\dot{\sigma}^{2}- V(\phi, \sigma),
\end{align}
and the corresponding EoS is now given by 
\begin{equation}
   w_\mathrm{Q} = \frac{p_\mathrm{Q}}{\rho_\mathrm{Q}}=\frac{\dot{\phi}^{2}-\dot{\sigma}^{2}-2V(\phi,\sigma)}{ \dot{\phi}^{2}-\dot{\sigma}^{2}+2V(\phi,\sigma)}.
\end{equation} 

Generally, there are two basic types of quintom models \cite{Guo:2006pc}. One is quintom-A, where the dark energy EoS parameter can cross the cosmological boundary $-1$ from quintessence regime to phantom regime as the universe expands, and the other is quintom-B type for which the EoS is arranged to change from below $-1$ to above $-1$.
In addition to the quintom model of two scalar fields, oscillating quintom model \cite{Feng:2004ff}, spinor fields \cite{Alimohammadi:2008mh}, string theory \cite{Cai:2007gs}, DHOST \cite{Langlois:2017mxy, Langlois:2018jdg} and Horndeski theory \cite{Horndeski:1974wa} can also be considered for the model-building. Quintom dynamics can also be
utilized to realize cyclic cosmology \cite{Cai:2007qw, Novello:2008ra}.

\begin{figure}[htbp]
    \centering
	\includegraphics[width=0.5\textwidth]{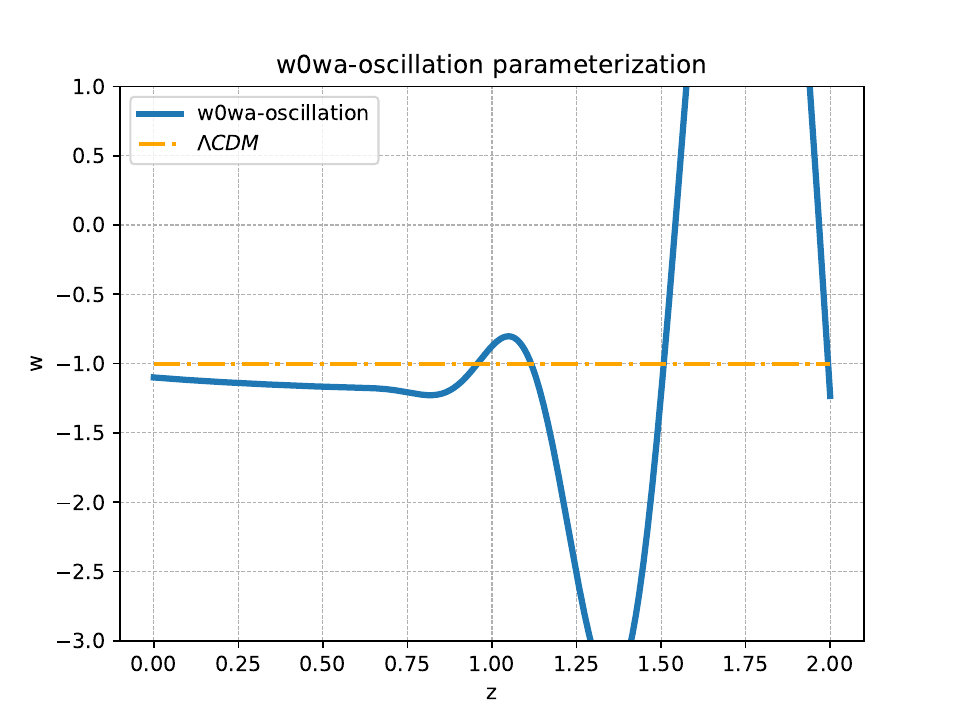}
	\caption{{\it{An example of oscillating dark energy realization
    within  the $w_0w_a$-oscillation parametrization. The parameter values have been chosen as $w_0=-1.1$, $w_a=-0.2$, $w_\mathrm{os}=-8.0$, $\Omega_{\mathrm{os}}=15.0\pi$, $\beta=80.0$ and $a_*=0.3$ (all   parameters are dimensionless).}}}
	\label{fig:wde oscillating}
\end{figure}

In order to provide an example  we consider a dark energy EoS parameter   parametrized as
\begin{equation}
    w (a) = w_0+w_a(1-a)+w_\mathrm{os} \sin{(\Omega_{\mathrm{os}} a)} e^{-\beta(a-a_*)^2}.
\end{equation}
This parameterization adds an extra oscillation term to the traditional $w_0w_a$ parametrization, and thus we call it the $w_0w_a$-oscillation parameterization. $w_0$ and $w_a$ have the same interpretation
as in the $w_0w_a$ parameterization: $w = w_0$ at the current time and
$w = w_0 + w_a$ when $a = 0$. For the additional oscillating part, $w_{\mathrm{os}}$ represents the amplitude of the oscillating part, $\Omega_{\mathrm{os}}$ represents the angular frequency of the oscillation, $\beta$ represents the rate of decay of the oscillation and $a_*$ represents the moment at which the oscillation begins.
Since we require $w(a=1)=w_0$, we have that $\sin{(\Omega_{\mathrm{os}})}=0$, leading to $\Omega_{\mathrm{os}}=n \pi$, where $n$ is a non-zero integer.
Using this parametrization, the evolution of the dark energy EoS
parameter is shown in Fig.~\ref{fig:wde oscillating}.

\subsection{The effective field theory of dark energy}

The effective field theory (EFT) of dark energy represents a comprehensive and versatile framework for studying all single-field dark energy models\cite{Creminelli:2008wc, Park:2010cw, Gleyzes:2013ooa}, as well as curvature-based modifications of gravity \cite{Gubitosi:2012hu, Bloomfield:2012ff, Frusciante:2019xia}. 
This theory is capable of describing both the evolution of the cosmological background and the perturbations that arise within it. One of the key strengths of the EFT approach is its ability to decouple the treatment of perturbations from that of the background, allowing each one to be analyzed independently. These features allow the evolution behavior of the three kinds of dark energy to be described in the framework of EFT.

In the EFT framework one usually adopts the unitary gauge, in which the scalar field $\phi$ does not explicitly appear in the EFT action. Instead, the scalar degree of freedom is absorbed into the metric.
More precisely, the time coordinate is defined as a function of the scalar field, leading to the vanishing of fluctuations around the background: \(\delta \phi(t,\Vec{x}) \equiv \phi(t,\Vec{x}) - \phi_0(t) = 0\), where \(\phi_0(t)\) denotes the background value of the scalar field. 
To reintroduce the scalar field degree of freedom alongside full diffeomorphism invariance, one may employ the ``Stueckelberg trick''. 
This involves performing an infinitesimal-time diffeomorphism \(t \rightarrow t + \pi(x)\), where \(\pi(x)\) serves as the field perturbation that encapsulates the scalar dynamics of dark energy. 

Through this method, the scalar field role in the EFT framework can be restored, providing a more complete picture of the system under study. The general EFT action with unitary gauge can be written as
\begin{equation}
    S=\int \mathrm{d}^4x \sqrt{-g}\left[ \frac{M_{\mathrm{P}}^2}{2}\Psi (t)R-\Lambda(t)-c(t)g^{00}\right]+S_{\mathrm{de}}^{(2)},
    \label{eq:eft action}
\end{equation}
with $M_{\mathrm{P}}^2=1/8\pi G$   the Planck mass, and where $\Psi$, $\Lambda$ and $c$ are functions of the time coordinate $t$. Additionally, 
$S_{\mathrm{de}}^{(2)}$ indicates terms that start explicitly quadratic in the perturbations and therefore do not affect the background (for  more details on the EFT perturbation see \cite{Pearson:2012kb}). Hence, different dark energy models will yield different expressions of those parameters. 

In order to extract the background evolution, we   vary the above action with respect to the metric, obtaining the modified Friedmann equations in a flat universe, namely
\begin{align}
    3H^2&=\frac{1}{M_{\mathrm{P}}^2 \Psi}\Big (\rho_{\mathrm{m}}-3M_{\mathrm{P}}^2H\dot{\Psi}+c+\Lambda \Big ) \nonumber \\
    &=\frac{1}{M_{\mathrm{P}}^2 \Psi}\Big (\rho_{\mathrm{m}}+\rho_{\mathrm{de}} \Big )=\frac{1}{M_{\mathrm{P}}^2}\Big (\rho_{\mathrm{m}}+\rho_{\mathrm{de}}^{\mathrm{eff}} \Big ) \\
    -2\dot{H}-3H^2&=\frac{1}{M_{\mathrm{P}}^2 \Psi}\Big (p_{\mathrm{m}}+M_{\mathrm{P}}^2\Ddot{\Psi}+2M_{\mathrm{P}}^2H\dot{\Psi}+c-\Lambda \Big )  \nonumber \\
    &=\frac{1}{M_{\mathrm{P}}^2 \Psi}\Big (p_{\mathrm{m}}+p_{\mathrm{de}} \Big )=\frac{1}{M_{\mathrm{P}}^2}\Big (p_{\mathrm{m}}+p_{\mathrm{de}}^{\mathrm{eff}} \Big ).
\end{align}
From the above equations  we can find that $\Psi$ indicates whether the scalar field is minimally coupled. One can thus define the density and pressure of dark energy as
\begin{align}
    \rho_{\mathrm{de}}^{\mathrm{eff}}&=\frac{1-\Psi}{\Psi}\rho_{\mathrm{m}}-3M_{\mathrm{P}}^2H\frac{\dot{\Psi}}{\Psi}+\frac{c}{\Psi}+\frac{\Lambda}{\Psi} \nonumber \\
    &=\frac{1-\Psi}{\Psi}\rho_{\mathrm{m}}+\frac{\rho_{\mathrm{de}}}{\Psi}, \\
    p_{\mathrm{de}}^{\mathrm{eff}}&=\frac{1-\Psi}{\Psi}p_{\mathrm{m}}+M_{\mathrm{P}}^2\frac{\Ddot{\Psi}}{\Psi}+2M_{\mathrm{P}}^2H\frac{\dot{\Psi}}{\Psi}+\frac{c}{\Psi}-\frac{\Lambda}{\Psi} \nonumber \\
    &=\frac{1-\Psi}{\Psi}p_{\mathrm{m}}+\frac{p_{\mathrm{de}}}{\Psi}.
\end{align}
Thus,  the effective dark energy EoS parameter for a general EFT model can be written as
\begin{equation}
    \begin{aligned}
    w_{\mathrm{EFT}}^{\mathrm{eff}}&=\frac{(1-\Psi)p_{\mathrm{m}}+M_{\mathrm{P}}^2\Ddot{\Psi}+2M_{\mathrm{P}}^2H\dot{\Psi}+c-\Lambda}{(1-\Psi)\rho_{\mathrm{m}}-3M_{\mathrm{P}}^2H\dot{\Psi}+c+\Lambda} \\
    &=-1+\frac{(1-\Psi)(\rho_{\mathrm{m}}+p_{\mathrm{m}})+M_{\mathrm{P}}^2\Ddot{\Psi}-M_{\mathrm{P}}^2H\dot{\Psi}+2c}{(1-\Psi)\rho_{\mathrm{m}}-3M_{\mathrm{P}}^2H\dot{\Psi}+c+\Lambda}.
    \end{aligned}
\end{equation}

We proceed by  presenting   examples on how to acquire those EFT parameters from   specific models. 
For $\Psi (t)=1$  the action \eqref{eq:eft action} contains any minimally-coupled single-field dark energy model. 
If we consider   standard quintessence scenario, we should first rewrite the Lagrangian in the unitary gauge as
\begin{equation}
    -\frac{1}{2}(\partial \phi)^2-V(\phi) \stackrel{\mathrm{unitary}}{\longrightarrow}-\frac{1}{2}\dot{\phi}_0^2(t)g^{00}-V(\phi_0),
\end{equation}
which should correspond to the background action of \eqref{eq:eft action}.
Thus, we obtain
\begin{equation}
    \Psi (t)=1, \quad c(t)=\frac{1}{2}\dot{\phi}_0^2(t), \quad \Lambda(t)=V(\phi_0).
\end{equation}

In the case of $f(R)$ gravity we can rewrite the action in the unitary gauge by choosing the background value $R^{(0)}=t$ as
\begin{equation}
    f(R) \stackrel{\mathrm{unitary}}{\longrightarrow} f_R(R^{(0)})R+f(R^{(0)})-R^{(0)}f_R(R^{(0)}),
\end{equation}
and thus we obtain
\begin{equation}
\begin{aligned}
    \Psi (t)&=f_R(R^{(0)}), \quad c(t)=0, \\
    \Lambda(t)&=-\frac{M_{\mathrm{P}}^2}{2}\Big (f(R^{(0)})-R^{(0)}f_R(R^{(0)}) \Big ).
\end{aligned}
\end{equation}

Moreover, proceeding beyond the curvature-based EFT theory, the torsion-based EFT was established in the works \cite{Li:2018ixg, Cai:2018rzd}, and the torsion-based EFT action in the unitary gauge is given by
\begin{equation}
\begin{aligned}
    S=&\int \mathrm{d}^4x \sqrt{-g}\Bigg [ \frac{M_{\mathrm{P}}^2}{2}\Psi (t) \mathring{R}-\Lambda(t)-c(t)g^{00}  + \frac{M_{\mathrm{P}}^2}{2}d(t)T^{0} \Bigg ]+S_{\mathrm{de}}^{(2)},
    \label{eq:EFT f(T)}
\end{aligned}
\end{equation}
where $\mathring{R}$ represents the Ricci scalar calculated by the Levi-Civita connection, while $T^0=T^{\mu}_{\ 0\mu}$.
Notice that if the term $T^{0}$ vanishes, the above formalism is the same as the curvature-based EFT theory. Specifically, for $f(T)$ theory  we fix the time slicing in order to coincide with the uniform $T$ hypersurfaces, since doing so would enforce the terms in the expansion of $f(T)$ action around the background value $T^{(0)}$ beyond the linear   order to vanish, since their contribution to the equations of motion would always include at least one power of $\delta T$.
Thus, in the unitary gauge we have
\begin{equation}
    f(T) \stackrel{\mathrm{unitary}}{\longrightarrow}  f_T(T^{(0)})T+f(T^{(0)})-f_T(T^{(0)})T^{(0)}.
\end{equation}
Recall the relation that the torsion scalar $T$ is different from $\mathring{R}$ by a boundary term, namely
\begin{equation}
    \mathring{R}=-T-2\nabla_{\mu}T^{\mu},
\end{equation}
where $T^{\mu}=T^{\nu\mu}_{\ \ \ \nu}$. After integrating by parts and dropping the boundary term, we rewrite the action of $f(T)$ gravity under the unitary gauge as
\begin{equation}
    \begin{aligned}
    f(T) \stackrel{\mathrm{unitary}}{\longrightarrow} & -f_T(T^{(0)}) \mathring{R}+2\dot{f_T}(T^{(0)})T^{(0)}-T^{(0)}f_T(T^{(0)})+f(T^{(0)}).
    \end{aligned}
\end{equation}
Compared with Eq.~\eqref{eq:EFT f(T)}, we can observe that
\begin{equation}
    \begin{aligned}
        \Psi(t)&=-f_T(T^{(0)}), \quad c(t)=0, \quad d(t)=2\dot{f_T}(T^{(0)}) \\
        \Lambda(t)&=-\frac{M_{\mathrm{P}}^2}{2}\Big (f(T^{(0)})-T^{(0)}f_T(T^{(0)}) \Big ).
    \end{aligned}
\end{equation}
Hence, one is able to present an appropriate realization of the phenomena appeared in the previous section by virtue of the torsional based EFT. Particularly, this approach has been developed in the literature to address other cosmological scenarios \cite{Yan:2019gbw, Yan:2019hxx, Cai:2008gk}.

\section{Conclusions}\label{conclusion}

Since the release of the first-year BAO observations from DESI, dynamical dark energy has attracted increasing attention. In this paper, we summarized the propensity of different cosmological observations for dynamical dark energy and we provided a brief review about the perspective from phenomenology and theoretical mechanisms.
We used a combination of different observations at the background level,   including SNe, BAO and CC ones, in order to reconstruct the evolution of the  dark energy EoS parameter by performing Gaussian process regression.
Additionally, we added the perturbation observations of matter growth rate RSD to reconstruct the deviation from GR for prudent model selection.
Moreover, we summarized the behavior from different combinations and we presented examples for theoretical implementations. 

Firstly, we considered the background-level observation data. For all reconstruction results  we found that in the redshift range less than $0.6$  we obtained a  significant agreement with the standard $\Lambda$CDM paradigm. 
Although the results of the SNe reconstruction   slightly deviate from the $\Lambda$CDM scenario, they are still within the range of 2$\sigma$.
After considering the BAO data too, we found   consistency between the CC   and BAO data, that is  the incorporation of BAO does not significantly affect
the evolution behavior of the result.
However, the joint analysis of SNe and BAO revealed that in the redshift range greater than $1.5$, $\Lambda$CDM scenario would exceed the 2$\sigma$ range shown by the reconstruction of $X$, which indicates that there may be an
inconsistency between SNe data and BAO data at high redshifts,
or some signals for dynamical dark energy with quintom-like behavior. 
We mention that the dynamical evolution of dark energy does share some common characteristics,
and we divided these characteristics into three main different categories, which are negative-energy dark energy, late-dominated dark energy and oscillating dark energy. 
Furthermore, after considering the effect of modified gravity on the growth of matter perturbations, we added to our analysis the RSD data from LSS. The results showed no significant departure from GR, hence only minor modifications can be imposed.
 
Then, for each of these different types of dark energy dynamics, we provided   concrete examples. For the negative-energy dark energy model  we tried to explain it using modified gravity, and we presented an example of this kind of dynamical evolution within $f(T)$ gravity theory. 
For late-dominated dark energy, the single scalar field theory was taken into consideration, especially the quintessence scenario. 
For the oscillating dark energy  we parameterized the evolution by adding a damped oscillation term to the traditional $w_0w_a$ parametrization. Finally, we showed that all these different theories can be unified in the framework of EFT.

In summary, on   one hand  we expect that in the future there will be     cosmological observations of increasing accuracy, which may constantly refresh our understanding of the nature of the universe. On the other hand, it is     necessary to advance the theory of dark energy too. Although the microphysical nature of dark energy remains a mystery, the latest cosmological observations indicate that we may be on the eve of discovering the details of the dynamics of dark energy.

\section*{Acknowledgments}
We are grateful to Zhiyu Lu, Dongdong Zhang, Xinmin Zhang and Teruaki Suyama for insightful comments. This work was supported in part by the National Key R\&D Program of China (2021YFC2203100, 2024YFC2207500), by the National Natural Science Foundation of China (12433002, 12261131497, 92476203), by CAS young interdisciplinary innovation team (JCTD-2022-20), by 111 Project (B23042), by Anhui Postdoctoral Scientific Research Program
Foundation (No. 2025C1184), by CSC Innovation Talent Funds, by USTC Fellowship for International Cooperation, and by USTC Research Funds of the Double First-Class Initiative. 
ENS acknowledges the contribution of the LISA CosWG and the COST Actions  and 
of COST Actions CA21136 ``Addressing observational tensions in cosmology with 
systematics and fundamental physics (CosmoVerse)'',  CA21106 ``COSMIC WISPers 
in the Dark Universe: Theory, astrophysics and experiments (CosmicWISPers)'', 
and CA23130 ``Bridging high and low energies in search of quantum gravity 
(BridgeQG)''. Kavli IPMU is supported by World Premier International Research Center Initiative (WPI), MEXT, Japan.

\bibliographystyle{JHEP}
\bibliography{DESI}
\end{document}